\documentclass[12pt,tightenlines,aps,prb,superscriptaddress]{revtex4}

\usepackage{times}
\usepackage{color}

\definecolor{mypink}{RGB}{219, 48, 122}
\definecolor{mygreen}{RGB}{51, 153, 102}
\definecolor{turq}{RGB}{64, 224, 208}
\definecolor{aquam}{RGB}{19,179,172}
\definecolor{brown}{RGB}{165, 42, 42}
\definecolor{orange}{RGB}{253, 106, 2}

\newcommand{\GMT}{Ge$_{1-x}$Mn$_x$Te}
\newcommand{\CdMT}{Cd$_{1-x}$Mn$_x$Te}
\newcommand{\GMTD}{Ge$_{0.87}$Mn$_{0.13}$Te}
\newcommand{\lll}{$\langle 111 \rangle$}

\newcommand{\aGeTe}{\hbox{$\alpha$-GeTe}}
\newcommand{\Mni}{$\mathrm{Mn_i}$}
\newcommand{\Mns}{$\mathrm{Mn_s}$}
\newcommand{\MnL}{$\mathrm{Mn_{L3}}$}
\newcommand{\OM}{$\mathrm{m_{L}}$}
\newcommand{\SM}{$\mathrm{m_{S}}$}
\newcommand\degC{$\,^{\circ}\mathrm{C}$}

\usepackage{graphicx}

\begin{document}
\title{Collective topological spin dynamics in a correlated spin glass} 

\author{Juraj Krempask{\'y}}
\email{juraj.krempasky@psi.ch}
\affiliation{\footnotesize{Photon Science Division, Paul Scherrer Institut, CH-5232 Villigen, Switzerland}}

\author{Gunther Springholz}
\affiliation{\footnotesize{\hbox{Institut f{\"u}r Halbleiter-und Festk{\"o}rperphysik, Johannes Kepler Universit{\"a}t, A-4040 Linz, Austria}}}

\author{Sunil~Wilfred~D'Souza}
\affiliation{\footnotesize{New Technologies-Research Center University of West Bohemia, Plze{\v n}, Czech Republic}}

\author{Ond\v rej~Caha}
\affiliation{\footnotesize{\hbox{Dept. of Condensed Matter Physics, Masaryk University, Kotl\'a\v rsk\'a 267/2, 61137 Brno ,Czech Republic}}}

\author{Martin~Gmitra}
\affiliation{\footnotesize{\hbox{Institute of Physics, P. J. {\v S}af{\'a}rik University in Ko{\v s}ice, Park Angelinum 9, 040 01 Ko{\v s}ice, Slovakia}}}

\author{Andreas~Ney}
\affiliation{\footnotesize{\hbox{Institut f{\"u}r Halbleiter-und Festk{\"o}rperphysik, Johannes Kepler Universit{\"a}t, A-4040 Linz, Austria}}}

\author{Carlos~Antonio~Fernandes~Vaz}
\affiliation{\footnotesize{Photon Science Division, Paul Scherrer Institut, CH-5232 Villigen, Switzerland}}

\author{Cinthia~Piamonteze}
\affiliation{\footnotesize{Photon Science Division, Paul Scherrer Institut, CH-5232 Villigen, Switzerland}}

\author{Mauro~Fanciulli}
\affiliation{\footnotesize{LPMS, Universit\'e de Cergy-Pontoise, 95031 Cergy-Pontoise, France}}

\author{Dominik~Kriegner}
\affiliation{\footnotesize{Institute of Physics ASCR, v.v.i., Cukrovarnick{\'a} 10, 162 53 Praha 6, Czech Republic}}
\affiliation{\footnotesize{\hbox{Dept. of Condensed Matter Physics, Charles University, Ke Karlovu 5, 121 16 Praha 2, Czech Republic}}}

\author{Jonas~A.~Krieger}
\altaffiliation[Current address: ]{Max Planck Institut f\"ur Mikrostrukturphysik, Weinberg 2, 06120 Halle, Germany}
\affiliation{\footnotesize{\hbox{Laboratory for Muon Spin Spectroscopy, Paul Scherrer Institute, CH-5232 Villigen PSI, Switzerland}}}

\author{Thomas~Prokscha}
\affiliation{\footnotesize{\hbox{Laboratory for Muon Spin Spectroscopy, Paul Scherrer Institute, CH-5232 Villigen PSI, Switzerland}}}

\author{Zaher~Salman}
\affiliation{\footnotesize{\hbox{Laboratory for Muon Spin Spectroscopy, Paul Scherrer Institute, CH-5232 Villigen PSI, Switzerland}}}

\author{Jan~Min{\'a}r}
\email{jminar@ntc.zcu.cz}
\affiliation{\footnotesize{New Technologies-Research Center University of West Bohemia, Plze{\v n}, Czech Republic}}

\author{J.~Hugo~Dil}
\email{hugo.dil@epfl.ch}
\affiliation{\footnotesize{Photon Science Division, Paul Scherrer Institut, CH-5232 Villigen, Switzerland}}
\affiliation{\footnotesize{Institut de Physique, \'{E}cole Polytechnique F\'{e}d\'{e}rale de Lausanne, CH-1015 Lausanne, Switzerland}}

\begin{abstract}
\textbf{The interplay between spin-orbit interaction (SOI) and magnetic order is currently one of the most active research fields in condensed matter physics and leading the search for materials with novel and tunable magnetic and spin properties. Here we report on a variety of unexpected and unique observations in thin multiferroic \GMT\ films. The ferrimagnetic order in this ferroelectric semiconductor is found to reverse with current pulses six orders of magnitude lower as for typical spin-orbit torque systems. Upon a switching event, the magnetic order spreads coherently and collectively over macroscopic distances through a correlated spin-glass state. Lastly, we present a novel methodology to controllably harness this stochastic magnetization dynamics, allowing us to detect spatiotemporal nucleation of topological spin textures we term \hbox{``skyrmiverres''}}.
\end{abstract}

\maketitle

Recently, combining semiconducting and magnetic properties within the same material moved from being a concept  \cite{DMS_book} to being one of the main spintronic progress vectors \cite{Jungwirth_RMP_2014}, in particular when combining magnetism with topological properties \cite{Dietl_RMP_2014}. With the advent of ferroelectric Rashba semiconductors (FERS)  \cite{Picozzi_Front}, magnetic doping in dilute magnetic semiconductors (DMS) has opened pathways to exploit the electron spin associated with the Rashba-Zeeman type spin splitting of the electronic states \cite{Fabian_RMP_2004,JK_GMT,Yoshimi_GMT_2018}. To date the primary FERS representative is \aGeTe; with only two atoms per unit cell and  $\approx$0.3\ \AA\ displacement along the \lll\   direction between the Ge and Te atoms, it is arguably the simplest room temperature ferroelectric semiconductor\cite{Pawley_1966,JK_PRB}. The combination of the ferroelectric order and large SOI yields a switchable Rashba-type spin structure of the bulk states \cite{JK_PRX}, whereby the states are fully spin polarised around the valence band maximum \cite{JK_PRR_2020}. When doped with Mn a magnetic order is induced while the ferroelectric order remains at temperatures up to 250~K for dopings below 25\%, rendering it a multiferroic semiconductor \cite{Springholz_PRL, Kriegner_2016, JK_GMT}. Moreover, the collinear alignment of the magnetisation and FE polarisation axes ensures magnetoelectric coupling in the system \cite{JK_PRX}.
Conversely, as there is now a cohesive picture between Berry curvature and emergence of anomalous Hall effect in non-centrosymmetric magnetic semiconductors \cite{Nagaosa_AHE_RMP_2010}, the current-induced magnetization switching mechanism in this material points to a bulk Rashba-Edelstein effect \cite{Yoshimi_GMT_2018}. Both spectroscopic \cite{JK_GMT, JK_PRX} and transport studies \cite{Yoshimi_GMT_2018} indicate that  the key ingredient for \GMT\ magnetoelectric functionality is its itinerant magnetism mediated by the valence band \cite{Fukuma_2005,Sato_JPRPh_2005, Fukuma_2006, Fukuma_2008, Fukushima_2014}. Furthermore, this material is proposed to be a unique platform to explore novel phenomena such as nonreciprocal electric transport \cite{Yoshimi_PRB} or magnetic skyrmionic polarons \cite{Brey_Nano_2018}.

By combining a range of spectroscopic techniques based on X-ray magnetic circular dichroism (XMCD), near-edge x-ray absorption fluorescence spectroscopy (NEXAFS) and state of the art modeling (see Methods \ref{M_sprep},\ref{M_expm},\ref{M_comp},\ref{M_Ham}), we show that \GMTD\ is a bulk ferrimagnet (FiM). These results are further expanded upon with complementary low-energy muon spectroscopy (LE-$\mu$SR) as a local probe to characterize temperature-dependent magnetic transitions on nanometers scale; and SQUID magnetometry used to investigate the macroscopic magnetisation. The results testify that the FiM order builds up below $T_{c1}=100$ K on top of a paramagnetic (PM) background and that below $T_{c2}=40$ K the system behaves as a spin-glass. Developing new analysis tools based on magnetostochastic resonance (MSR) switching allows us to characterize the \GMTD\ magnetism in static and dynamical regimes.  
The ensemble of experimental techniques and theory yields a holistic view of a correlated spin glass with emergent topological spin textures. Our data indicate that XMCD can detect these topological spin-textures, that magnetization switching can be driven by nanoampere pulsed currents using stochastic resonance, that this switching is collective due to glassiness, and that the magnetic switch can be turned on and off by small changes in pulse periodicity.    

\subsection*{Ferrimagnetic full switching}

\begin{figure}[ht]
    \includegraphics[width=\textwidth]{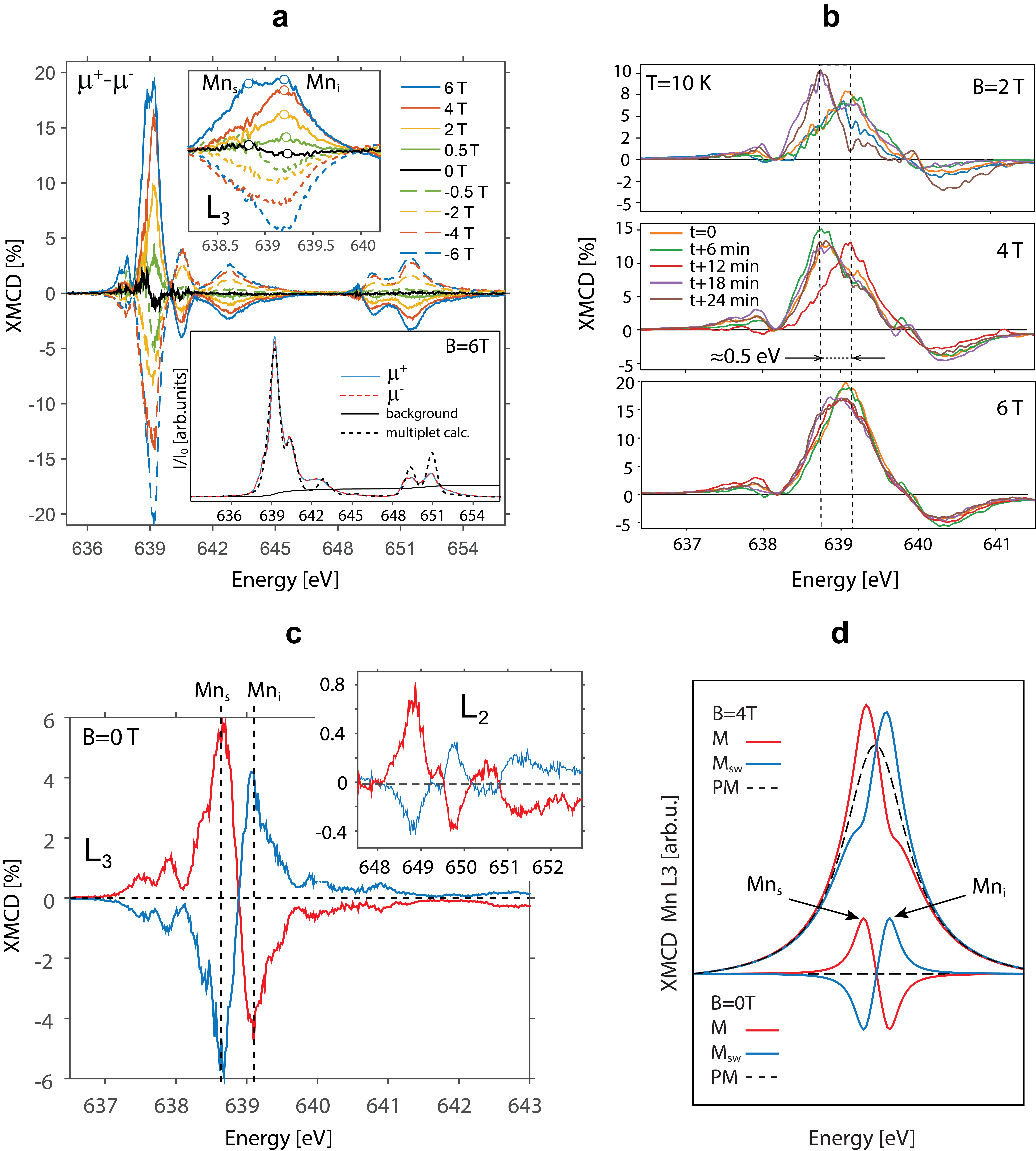}
    \caption{\textbf{Switching of ferrimagnetic order in XMCD.} (a) XMCD spectra for selected applied $B$-fields. (Top inset) zoom into the Mn $L_3$ edge, showing two distinct spectral weight from substitutional \Mns\ and interstitial \Mni\ atoms. (Bottom inset) background subtracted XAS spectra for $\pm \mu$ light polarizations, measured at B=6\ T (10\ K), and overlaid with multiplet calulations. (b) Series of \MnL\ XMCD spectra taken at $\approx$3~min intervals for selected \hbox{B-fields}. (c) Spontaneous switching recorded in two consecutive XMCD datasets (red/blue) in the absence of applied B-field, shown at the \MnL\ and $\mathrm{Mn_{L2}}$ absorption edges. d) \hbox{\MnL-edge} model based on 4 Gaussian functions simulating \Mns\ and \Mni\ spectral contributions. For PM the \Mns\ and \Mni\ features are aligned, with their amplitude dependent on the B-field. The FiM consists of oppositely aligned peaks with different amplitude. $M$ and $M_{SW}$ refer to the reference and switched state respectively. } 
    \label{F1}
\end{figure}

Figure~\ref{F1}a shows XMCD spectra measured at 10~K in total electron yield (TEY) mode while applying selected magnetic fields up to $\pm$~6 Tesla. As will be discussed later, the dichroism signal does not saturate at our maximum B-fields due to a paramagnetic spin-glass state, characteristic also for \CdMT\ related DMS system \cite{Galazka_1980}. The top-inset is a zoom-in into the \MnL\ absorption edge, showing two distinct XMCD features separated by $\approx$0.5\ eV. A time series of XMCD spectra for selected B-fields in Fig.~\ref{F1}b shows that these features are fluctuating, eventually completely and spontaneously reversing at zero B-field (Fig.~\ref{F1}c). The shape of the dichroism signal at the \MnL-edge can be understood from the combination of a paramagnetic background superposed with ferrimagnetic contributions (\ref{F1}d). The fluctuating spectral weight in Fig.~\ref{F1}b indicates that the switching continues for applied fields even above 4~T.  From considering an extensive data set obtained from several experimental runs and samples, the magnetic reversal appears to be regular and does not show the telegraph noise expected for a system which randomly switches between two magnetic, or spin, states \cite{Natterer_PRL_2018, Hermenau_NC2019}. As surprising as the switching itself, is the fact that it occurs simultaneously, within the time scale of the experiment, over at least the square millimetre area of the photon spot. 

To reiterate, the dichroism effect at \MnL\ systematically shows two opposite magnetic moments from different types of Mn atoms, with absorptions at $h\nu=$638.8 and 639.3~eV, simultaneously switching after about every energy scan across the \MnL\ absorption edge, in absence or presence of an externally applied field.  Notably, the opposite magnetic moments are not exactly equal in magnitude, indicating a non-balanced ferrimagnetic order from two distinct Mn-sites. The spontaneous switching is further corroborated by the fact that we obtain equivalent dichroism effect by measuring spectra with the same circular polarization  (see Extended Fig.~\ref{S1}a). This also indicates that the variation of light helicity is not the driving force of the switching. There are now three main question which arise:
\vspace{-0.5em}
\begin{itemize}
 \setlength\itemsep{-0.5em}
\item What is the difference between the two Mn-sites?
\item Why are their magnetic moments spontaneously reversing in time over macroscopic length scales?
\item How to control the switching?
\end{itemize}
\vspace{-0.5em}
We first identify the different Mn contributions with our theoretical calculations, and proceed with the magnetostochastic resonance switching model to address the latter two questions.         

\subsection*{Magnetic ground state properties}

In accordance with our earlier resonant ARPES studies \cite{JK_GMT} and extensive XMCD studies on other \hbox{Mn-doped} DMS systems \cite{GvL_PRB_GaMnN_2007,GvL_2014,GvL_GaMnAs_2015}, we identify these moments to originate from substitutional (\Mns) and interstitial (\Mni) atoms, respectively. However, in contrast to related systems we find that in \GMTD\ the characteristic \hbox{$\approx$0.5\ eV} energy separation between \Mns\ and \Mni\ is not related to Mn segregation effects, or to non-magnetic surface Mn-oxides, nor to a depletion of Mn in the near-surface layer. Instead, the energy shift is an intrinsic bulk property we also see in bulk-sensitive fluorescence yield and in NEXAFS analysis from the Mn-K edge (see Extended Fig.~\ref{S1}b-c, respectively). This confirms the presence of \Mns\ and \Mni\ atoms, with a \Mns:\Mni\ occupancy close to 2:1. We employ this experimental finding to calculate the \GMTD\ magnetic ground state properties using density functional theory, where the Mn-disorder was implemented within the coherent potential approximation (CPA) alloy theory (see Methods \ref{M_comp}).  

Figure \ref{F2} summarizes that Mn-doping induces a frustrated magnetism with various exchange interactions, while keeping the polar crystal structure. For 13\% Mn-doping we expect nearly 1 Mn per 8 unit cells, or 1 Mn atom randomly distributed over 4 unit cells in 2D. Based on the CPA formalism, Fig.~\ref{F2}a illustrates the Mn-occupancy in an ``infinite'' \GMTD\ lattice consisting of 91.1\% probability occupation of 8.9\% \Mns\ substituted Ge ($\mathrm{Mn_1}$), and two types of interstitial \Mni\ atoms $\mathrm{Mn_2}$ and $\mathrm{Mn_3}$, each one with 2.2\% probability occupation. The most prominent contribution to the magnetic order comes from the Heisenberg exchange energy between $\mathrm{Mn_{2,3}}$ and between $\mathrm{Mn_{1,2/3}}$ atoms, whereas the interaction between substitutional atoms ($\mathrm{Mn_{1,1}}$) becomes damped due to the presence of the interstitial Mn (Fig.~\ref{F2}b). The corresponding magnetic exchange couplings $J_{i,j}$ are presented as a function of the distance $R_{i,j}/a$ between atoms, where $a$ is the lattice parameter and $i,j=1,2,3$. The positive exchange constants ($J_{i,j}>0$) favours ferromagnetic (FM) order, whereas negative values ($J_{i,j}<0$) favour antiferromagnetic (AFM) order (see Methods \ref{M_Ham}).

The co-presence of FM/AFM in a DMS system are the main ingredients for magnetic frustration in a canonical spin glass, which in our case is further assited by Dzyaloshinskii-Moriya interaction (DMI). DMI is characteristic to non-centrosymmetric systems with large spin-orbit coupling (SOC) such as \GMT, where it promotes non-coplanar (canted) arrangement between spin states summarized in Fig.~\ref{F2}c. Both DMI and all $J_{i,j}$ have fluctuating $R_{i,j}/a$ dependence. Furthermore the DMI is non-negligible compared to the $J_{i,j}$, especially for larger distances, indicating significant magnetic frustration. For small length scales there is a competition between FM and AFM contributions, the relative magnitudes of which explain why \GMTD\ stabilizes in a ferrimagnetic ground state. Since the CPA approach allows us to address spectral contributions from individual Mn-atoms (Fig.~\ref{F2}d), the resulting theoretical magnetic dichroism can reproduce the measured spectra very well if we take into account that the majority of \Mns\ contribute to the PM background once the magnetic interactions stabilize, as detailed in Extended Fig.~\ref{S2}d. 

The here described FiM order, based on three distinct interacting Mn lattice sites as the ground state of the system, was elusive in previous theoretical \cite{Antonov_GMT_CMP_2010, Fukushima_2014} and experimental studies \cite{Fukuma_2005, Fukuma_2006}. The main reason for this is that in these studies only substitutional \Mns-sites were considered (empty markers in Fig.~\ref{F2}b), whereas in our case their oscillatory long-range $J_{i,j}$ exchange is damped due to the presence of interstitial \Mni\ atoms (filled green markers in Fig.~\ref{F2}b). Furthermore, when probed under applied B-field (Fig.~\ref{F1}b), the opposite FiM contributions are placed on a large PM background due to which the \Mni-moments partially align with those of \Mns. Only the dynamics shown in Fig.~\ref{F1}c allows to distinguish the two sites under applied magnetic field.

In order to assess the energetics of the FiM switching, Fig.~\ref{F3}a compares the energy scale of the rotation of \Mni\ spins with respect to the \Mns\ spin held fixed along the [111] axis with the magnetocrystalline anisotropy energy (MAE, see Methods \ref{M_Ham}). The combination of both energy profiles constitutes a double-well uniaxial potential with MAE barrier of $\approx$0.1\ meV comparable to ferromagnetic $3d$ metals \cite{MAE_PRL_95}. Having identified the general magnetic order and its bistable energy potential model, we now turn to the dynamics and the unexpected switching.

\subsection*{Magnetostochastic resonance switching}

Generally speaking, stochastic resonance (SR) is a phenomenon applicable to nonlinear systems whereby a weak signal is amplified through an entraining periodic signal or even noise. Besides the bistable potential, SR requires two additional ingredients \cite{Gammaitoni_RMP_1998}: (i) a weak periodic input and (ii) a source of noise entrained with the periodic input. As will be elaborated below, the first ingredient is provided by periodic TEY oscillations upon \GMTD\ x-ray illumination, whereas the second originates from the short TEY pulse in x-ray absorption during the scan over the \MnL\ absorption edge.  The sum of these assists the transition between the two FiM equilibria as depicted in Fig.~\ref{F3}a,b. 
 
The periodic drive $P(t)$ is in our case best described by a continuous sinusoidal signal with $T_\Omega\approx5$ minutes (Fig.~\ref{F3}c). These systematically observed oscillations are present also while performing h$\nu$ scans; they are not related to the periodic top-up current refill of the synchrotron and do not depend on x-ray polarization and applied B-field (see Extended Fig.~\ref{S3}a-c). Therefore, our understanding is that they originate from charging and discharging inside a capacitor circuit with \GMTD\ as dielectric, whereby the primary charging effect is caused by the ferroelectric order which under x-rays produce a charge separation due to a steady-state bulk photovoltaic effect \cite{vonBaltz_1981}. For the host GeTe this effect is even stronger compared to conventional oxide ferroelectrics \cite{Rappe_GeTeSHC_PRL_2018}. Moreover, the system possesses self-poling properties from host GeTe \cite{JK_MDPI_2019}, and negative capacitance \cite{Orlova_negGeTeC_arXiv}, leading to thermodynamically unstable transient effects \cite{Hoffmann_APL_2021}. Finally, the TEY current, especially as a short pulse $\eta(t)$ at the \MnL-edge (Fig.~\ref{F3}c), induces spin-orbit torque via the above mentioned Rashba-Edelstein effect \cite{Yoshimi_GMT_2018}.

A phenomenological description of magnetostochastic resonance (MSR) predicts a million times higher switching probability for bistable magnetic systems \cite{Grigorenko_IEEE_1995}. Consistent with these predictions, for \GMTD\ the current-driven magnetisation switching in the milliampere regime \cite{Yoshimi_GMT_2018} is achieved under MSR with a nanoampare TEY pulse at the \MnL\ absorption edge. This is implemented with \hbox{on-the-fly} absorption scans \cite{JK_OTF_AIP_2010}, in contrast to conventional slow \hbox{point-by-point} scans, where the FiM switching is not observed. A typical measurement series is denoted by the green top trace in Fig.~\ref{F3}d, where the resulting TEY signal is a two-frequency system in which $P(t)$ realizes frequency modulation and $\eta(t)$ implements amplitude modulation \cite{Song_AMFM_SR_2016}. With such a combination of driving frequencies the SR can be harnessed, e.g. controllably suppressed by detuning the frequencies from their optimal setting. We achieved this effect by measuring XMCD data with different energy ranges (grey bottom trace in \ref{F3}d), thereby moving away from the optimal synchronization with driving frequency $t_1=t_2=\frac{1}{2}t_{\Omega}$ \cite{Gammaitoni_RMP_1998} where $t_{1,2}$ is the time needed to measured one XAS spectrum, and $t_{\Omega}$ one XMCD dataset.

Figure \ref{F4} summarizes the MSR harnessing concept with two synchronized loops with aperiodic (MSR-off) and periodic additive noise (MSR-on). Data indicate that there is an immediate change from the MSR-off regime without FiM switching to FiM switching under the MSR-on regime. 

We emphasize that under MSR-on the FiM switching holds only in the statistical average, quantified via the residence time distribution $N(t)$ \cite{Gammaitoni_RMP_1998}. We quantified $N(t)$ by counting the multiples of $t_{\Omega}$ between subsequent FiM switching occurrences, leading to a sequence of Gaussian-like peaks with exponentially decreasing envelope as seen in Fig.~\ref{F4}d at 5\ K.  In qualitative agreement with SR predictions \cite{Gammaitoni_RMP_1998}, the peak distributions under thermal noise shifts to shorter times, as seen at 10\ K. To further confirm the SR origin of FiM switching, the $N(t)$ distribution envelope is found to depend on the resonance conditions through the driving period $t_{\Omega}$ (see Extended Fig.~\ref{S4}a). 

\subsection*{Topological spin textures}

In order to obtain a microscopic understanding of the MSR switching, we analyze the orbital \OM\ and spin \SM\ contribution from sum rules \cite{Thole_PRL_92,Carra_PRL_93}. Because the magnitude and sign of the MSR-on XMCD signal changes in time, the directional sense of the magnetic moments with respect to the beam axis is changing as well, occasionally measuring a totally quenched moment \hbox{$\bf{M}_{tot}=\mu_B(2\mathrm{m_{S}} + \mathrm{m_{L}})$}  such as in the grey traces in Fig.~\ref{F4}c. The B-field dependence of  \OM\ and  \SM\ contributions obtained for MSR-off and MSR-on regimes is summarized in Fig.~\ref{F5}.  For Mn the theoretically obtained \OM\ of -0.032$\mu_B$/(Mn atom) is independent of the magnitude of the applied field and the total magnetisation $\bf{M}_{tot}$ is expected to be dominated by \SM, which takes a value of  \SM$\approx$2.1$\mu_B$/(Mn atom) at saturation. This trend is well observed in Fig.~\ref{F5}a where in the MSR-off regime the applied B-field drives \SM\, whereas  \OM\ is rather constant. That the local Mn-moments deduced from the sum rules (0.17$\pm$0.01\ $\mu_B$/Mn atom) for B=0 differ from the theoretical values is due to the fact that most Mn-moments are canted away from the surface normal. Table~\ref{T2} in Methods \ref{M_comp} compares the calculated Mn-moments with experimental data under MSR-off/on regimes.

The MSR-on data in Fig.~\ref{F5}b,c show a similar picture with \SM\  following the magnetic field strength and \OM\ initially small and nearly constant. However, when the field direction is reversed, the \OM\ starts to strongly deviate from this constant. If the field continues to be rapidly changed, i.e. after each XMCD data pair, the \OM\ values return to the baseline (Fig.~\ref{F5}b). However, as shown in Fig.~\ref{F5}c, if the change in B is slower and the field steps smaller, the orbital moment \OM\ appears to ring up acquiring values that rise far above meaningful values for Mn atoms in our system. Moreover, the spin contribution \SM\ appears to be dragged along and increases to values expected for higher applied B-field, together leading to an exponentially increasing oscillatory total moment (Fig.~\ref{F5}c inset). This behavior and these unexpected high values strongly indicate a collective effect and imply a nucleation of topological spin textures (spin vortices) during the field sweeping, reminiscent to skyrmion nucleation \cite{dosSantos_NC_2016}.  The dependency on how fast the field is changed indicates that the ``spin viscosity'' changes, in correspondence with the spin glass character discussed below.

A macroscopic perspective on skyrmion nucleation in \GMTD\ is attested by SQUID magnetometry (Fig.~\ref{F5}d). First, the presence of topological spin textures in such films were anticipated from non-monotonic decrease (dips) in the magnetic field dependence of the anomalous Hall component \cite{Yoshimi_GMT_2018}. For bulk ferromagnets lacking inversion symmetry such dips are related to competition between the metastable skyrmions aggregation and magnetic domains, resulting in a non-equilibrium state under varying B-field \cite{Tokura_FeGe_NPhys_2018}. Because the remanent magnetization is lower for \GMTD\ samples grown on InP as for those grown on BaF$_2$, this re-aggregation effect is more pronounced in the former (Fig.~\ref{F5}d). Thus the dips indicated by arrows are indirect evidence of metastable skyrmion re-aggregation \cite{Tokura_FeGe_NPhys_2018}.  Finally, the unusual hysteresis shape with open area and narrow waist is characteristic to N\'eel-type skyrmions hosted in perpendicular magnetic thin-film ferromagnets \cite{Liu_ChPL_2022,Buttner_NM_2021}.  In our XMCD experiments the sudden \OM\ changes upon magnetic field reversal serve as a direct probe for the emergence of topological spin-textures, highlighting that XMCD can indeed detect unconventional magnetic structures such as skyrmions \cite{dosSantos_NC_2016}.

\subsection*{Correlated spin glass}

Already from the shape of the M(T) magnetization curve in Fig.~\ref{F6}b it is evident that the system does not behave like a ferromagnet with typical magnetisation $(1-(T/T_c)^{\alpha})^{\beta}$ dependence and a single transition temperature, but rather a steadily diminishing magnetization with various critical temperatures indicated in the figure. The splitting between zero-field-cooled (ZFC) and field-cooled (FC) curves is a clear indication of competing energy scales and spin orders, and of possible spin excitations in magnetic glassy-phases \cite{Vincent_SG_2018}. LE-$\mu$SR is the method of choice to characterize magnetically ordered volumes inside such a spin-glass from a local perspective. Fig.~\ref{F6}a summarizes three quantities extracted from our measurements: (i) the local mean field $B$ sensed by the implanted muons (top); (ii) the damping rate $\lambda$ of the oscillations which is proportional to the width of local field distribution (middle); and (iii) the initial amplitude of the precessing muon spin polarization $A_0$ (bottom). All three parameters start to change around the onset of magnetic order ($\approx$100~K) and show either a minimum ($B$), maximum ($\lambda$), or plateau ($A_0$) at 40~K.
The combination of bulk magnetization and LE-$\mu$SR data provides clear evidence that the magnetic order has a net magnetic moment imposed by the FiM order below $T_{c1}\approx$100~K, followed by a spin-glass transition below $T_{c2}\approx$40~K where the coupling between the FiM clusters becomes strong enough for the sample to undergo a transition to a cluster spin-glass long range order. 

The combined set of experimental and theoretical results allows us to rationalize the details of the MSR switching mechanism. For a broader perspective Fig.~\ref{F6}b compares the temperature dependence of $\mu$SR local mean field, bulk magnetization, and XMCD switching statistics. As expected, the spin-freezing temperature $T_{c2}$ is marking a minimum rather than an onset in FiM switching probability. Strikingly, as highlighted by the red area, above $T_{c2}$, the switching events dramatically increase along with the decreasing size of the FiM clusters; and this trend appears to be independent on whether the system was magnetized under ZFC or field-heated field-cooled (FH-FC) indicated by green and red markers, respectively. Moreover, at T=65~K the XMCD asymmetry sometimes reaches values as high as 60\% in contrast to the expected increase of magnetic disorder at elevated temperatures  (Extended Fig.~\ref{S4}). This is reminiscent of the spontaneous thermally induced magnetic order in elemental spin glasses \cite{Verlhac_NP_2022}, which in our case begins to morph between individual FiM clusters as indicated by pictographs in Extended Fig.~\ref{S4}b. We expect that such morphing/clustering create new spin vortex patterns which also affect the PM background, thereby enhancing the dichroism effect close to theoretical predictions in Fig.~\ref{F2}b. That kind of reordering on a mesoscopic scale is extremely slow, explaining the ``spin viscosity'' dynamical behavior seen in Fig.~\ref{F5}c inset. 

The slow, viscous, reordering on a mesoscopic scale (Fig.~\ref{F5}c inset) and the magnetic frustration highlighted in the theory section, are directly related to the glassy behavior. 
Furthermore, our XMCD results show that the FiM switching is collective over macroscopic dimensions. This indicates that the magnetic order has to be correlated across local magnetic interactions on which the magnetic spin textures spread like a collective excitation through the sample with spatiotemporal net spin windings generating the oscillatory magnetism as indicated in Fig.~\ref{F5}c inset. This is exactly what is expected for a correlated cluster spin-glass \cite{Ochoa_SHD_PRB, EManiv_collective_AFM_Science_2021} which under certain applied B-fields, attained after proper initialisation with B-field sweeping, can nucleate topological spin vortices on a mesoscopic scale. 

\section*{Conclusions}

Our \GMTD\  thin films present a variety of intriguing physical phenomena which we could decipher using our holistic approach; XMCD and theory confirm a bistable ferrimagnetic ground state, whereas the collective switching of this state in XMCD, together with $\mu$SR and magnetometry, indicate a cluster spin glass, and SQUID and XMCD show the emergence of topological spin textures supported by the large DMI found in theory. To highlight the novel concept of spatiotemporal topological spin textures arising from a spin glass, we name them ``skyrmiverres''. Given the rich phase diagram of most spin glasses with respect to variation of external parameters, we expect that such skyrmiverres under dynamical conditions are ubiquitous in multiferroic semiconductors such as \GMTD\ with metastable topological spin states coupled to magnetic and ferroelectric order. It will be of particular interest to determine the shape and distribution of these skyrmiverres using nanoscopic imaging techniques, especially in combination with pulsed currents to follow their switching behavior. For bistable systems we established a concept to harness magnetostochastic switching in XMCD experiments which allowed us to describe novel magnetization dynamics. Under this MSR drive the current needed for switching of the magnetic state is reduced by six orders of magnitude, allowing for the design of energy efficient non volatile FiM electronics possibly employing skyrmiverres as information carriers. Lastly, the combination of volatile and non-volatile elements and long and short term memory could make the system a promising playground for neuromorphic memory. On a more fundamental level, our results open up new avenues for studying collective topological spin dynamics in correlated spin glass systems.


\begin{thebibliography}{10}
\bibitem{statNote}
The FiM switching statistical error margins are binomial confidence intervals
  (obtaining $x$-switches in $n$ trials gives a $\sqrt(x/n*(1-x/n)/n)$ error).
  Statistically most relevant FiM switching summarized from three independent
  beamtimes from fresh samples was measured at 5~K (ZFC), that is why the
  smallest error bar in FiM switching probability in Fig.~6b is achieved under
  these measurement conditions.
\end{thebibliography}


\section*{Acknowledgments}
We thank Stefano Rusponi and Valerio Scagnoli for fruitful discussions. We gratefully acknowledge financial support from the Swiss National Science Foundation project No. PP00P2\_170591. Ja.M. and SdS would like to thank the CEDAMNF (Grant No. CZ.02.1.01/0.0/0.0/15\_003/0000358) co-funded by the Ministry of Education, Youth and Sports of Czech Republic and the GACR Project No. 20-18725S for funding. G.S. acknowledges support by the Austrian Science Funds (FWF), Projects P30960-N27 and I4493-N. DK acknowledges the  Lumina Quaeruntur fellowship of the Czech Academy of Sciences.
All experimental data will be made available upon request.

\newpage

\begin{figure}
    \includegraphics[width=\textwidth]{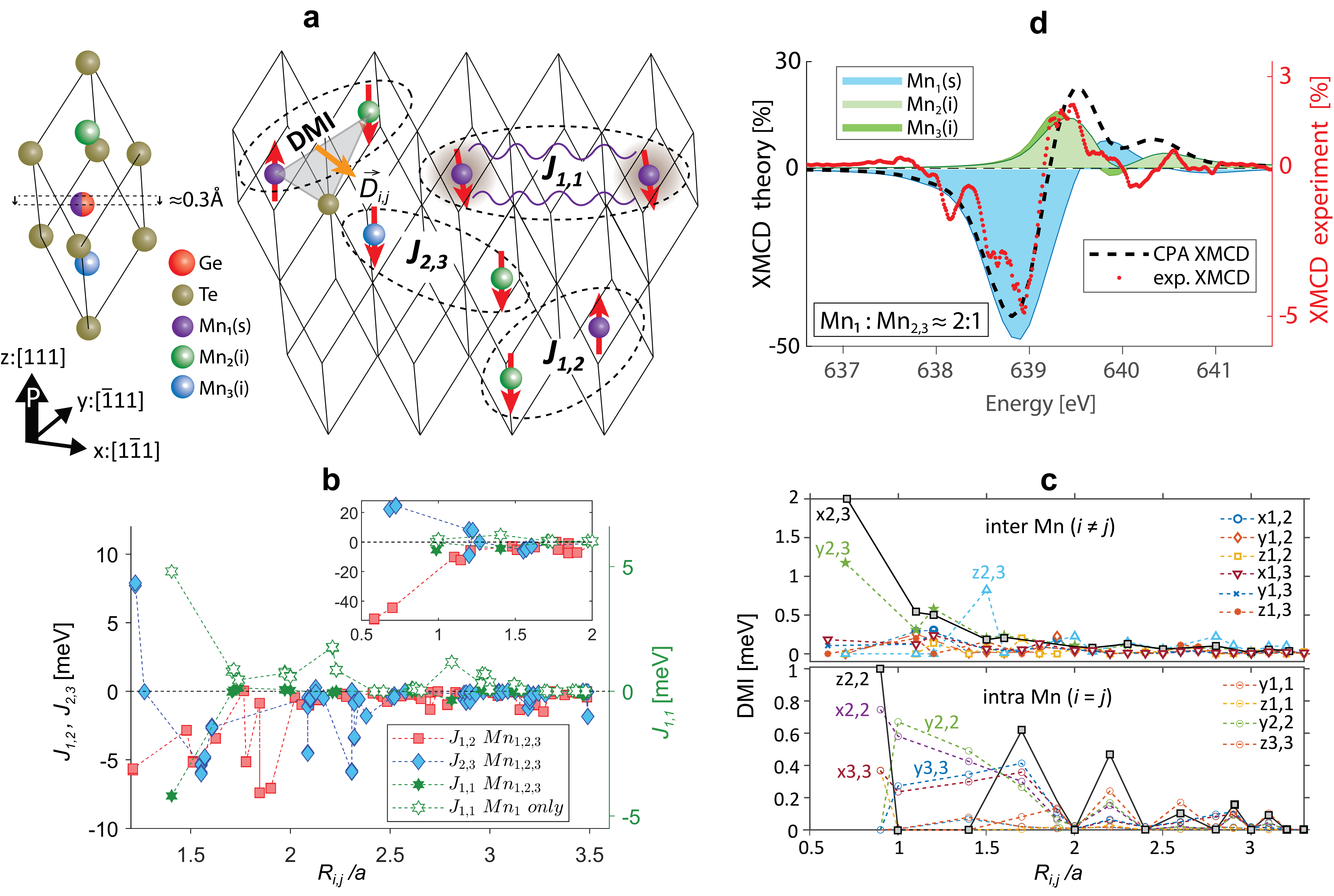}
    \caption{\textbf{Frustrated magnetic interactions in \GMTD.} (a) Illustration of Mn doping in the  rhombohedrally distorted \aGeTe\ unit cell with one substitutional \Mns\ and two potential interstitial \Mni\ dopants. (b) Exchange couplings $J_{i,j}$,  between \hbox{$\mathrm{Mn_i}$-$\mathrm{Mn_{j}}$} atoms with all three types of Mn atoms present (full markers), and with only substitutional $\mathrm{Mn_1}$ atoms (empty marker). Distances are normalized to the lattice parameter \textit{a}. (Inset) Extension to smaller atomic distances. (c) DMI values for inter- and intra-Mn site coupling. The $D_{i,j,x}$ component is summarized as $xi,j$ and similar for $y$ and $z$ ($x,y,z$ index directions are denoted in a). Dominant interactions are indicated in the graph, others listed in legends. (d) Comparison between XMCD experiment (red dotted line) and CPA theory, in which spectral contributions from individual Mn atoms result in a final XMCD signal (dashed line).}
    \label{F2}       
\end{figure}

\begin{figure}[ht]
    \includegraphics[width=\textwidth]{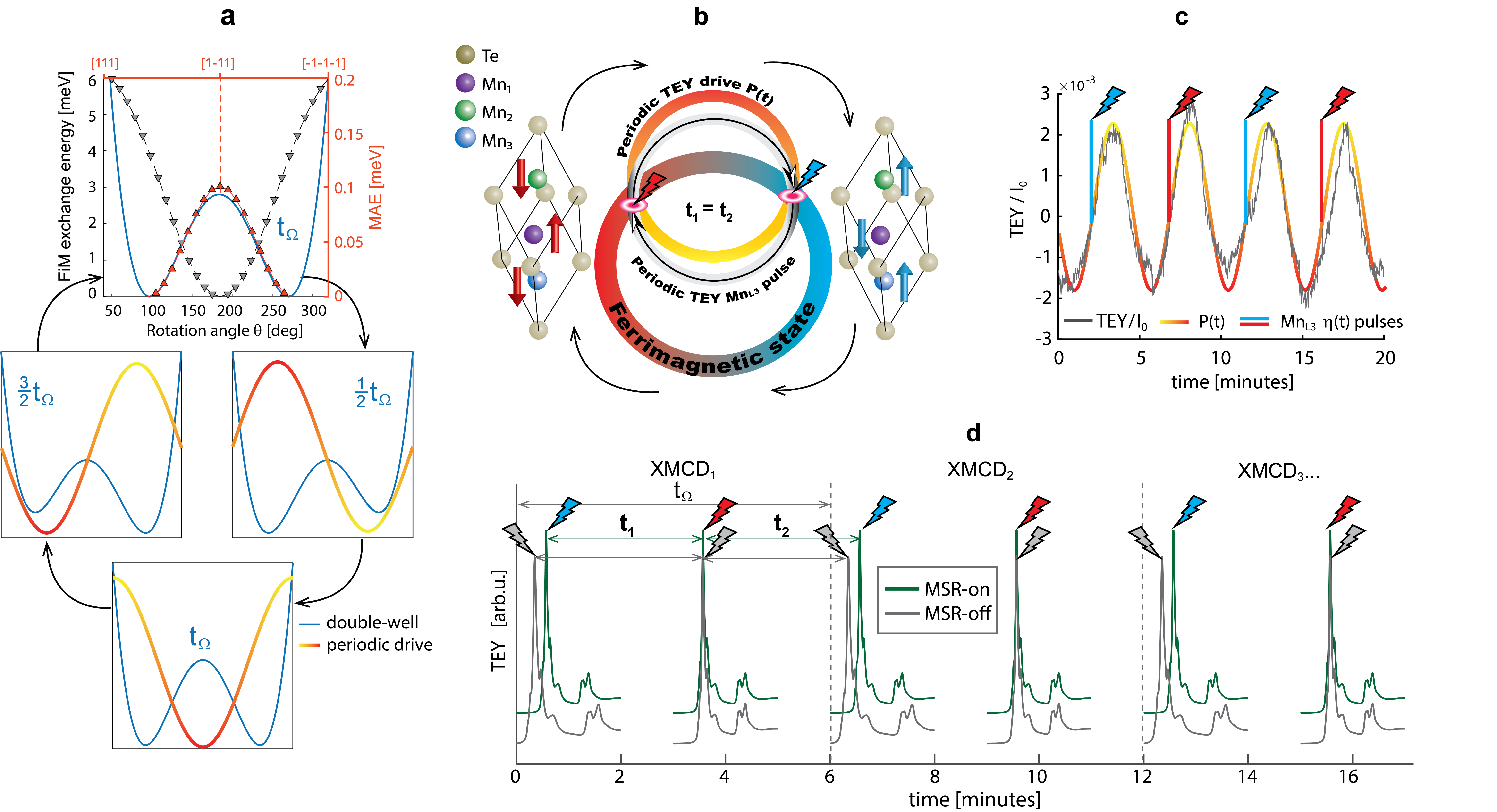}
    \caption{\textbf{Magnetostochastic resonance driving sources.} (a) Stochastic resonance switching on a symmetric double-well qualitatively modelled by FiM exchange energy (grey markers) and magnetocrystalline anisotropy energy (red markers). The potential is subject to a synchronised periodic drive $\mathrm{T_{\Omega}}$ (red-yellow trace), enabling the switching of the otherwise stable FiM state. (b) FiM switching cartoon: the circles illustrate the MSR harnessing concept with periodic \MnL\ TEY pulses indicated by grey and red/blue flash symbols, respectively. Periodic pulses with entrained $P(t)$ signal (red-yellow circle) enable the FiM switching in a statistical average.
(c) Sinusoidal  fit (red-yellow trace) of a time evolution of normalized TEY periodic drive $P(t)$ measured at \hbox{10\ K}, B=0, and $h\nu=638.8$eV (grey trace). Superimposed red and blue spikes represent transient TEY pulse at \MnL\ edge that would occur during XAS energy scans. (d) Typical time series of periodic (green/top) and aperiodic (grey/bottom) XAS scans which controls the magnetostochastic resonance switching (MSR-on/off). Empty spaces between the XAS scans is time needed to move the monochromator back to the initial energy.}
    \label{F3}

\end{figure}

\begin{figure}[ht]
    \includegraphics[width=\textwidth]{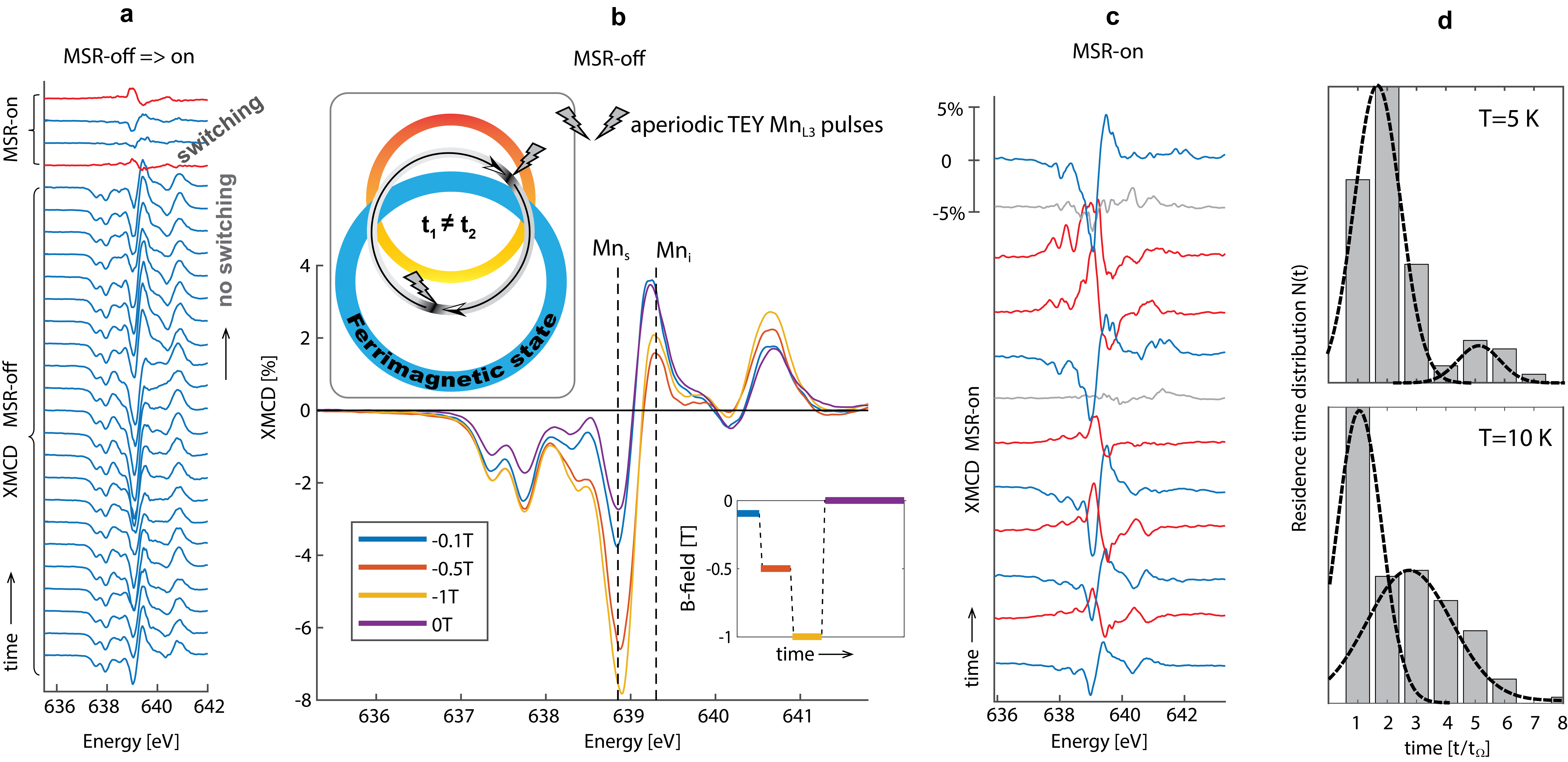}
    \caption{\textbf{Harnessing switching dynamics with magnetostochastic resonance.} (a) Stacked plot of \MnL\ XMCD spectra measured in \hbox{MSR-off} regime directly followed by MSR-on regime. (b) XMCD MSR-off spectra averaged for individual applied B-fields as depicted in the inset. Top cartoon illustrates that under aperiodic \MnL\ TEY pulses the MSR switching is disabled. (c) Stacked plot of MSR-on XMCD spectra. In (a) and (c) blue/red traces reflect the original or switched FiM state, grey traces reflect totally quenched magnetic state. (d) Statistical FiM switching quantification by residence time distributions $N(t)$, fitted with Gaussian functions at 5 and 10\ K.}
    \label{F4}
\end{figure}

\begin{figure}[ht]
    \includegraphics[width=\textwidth]{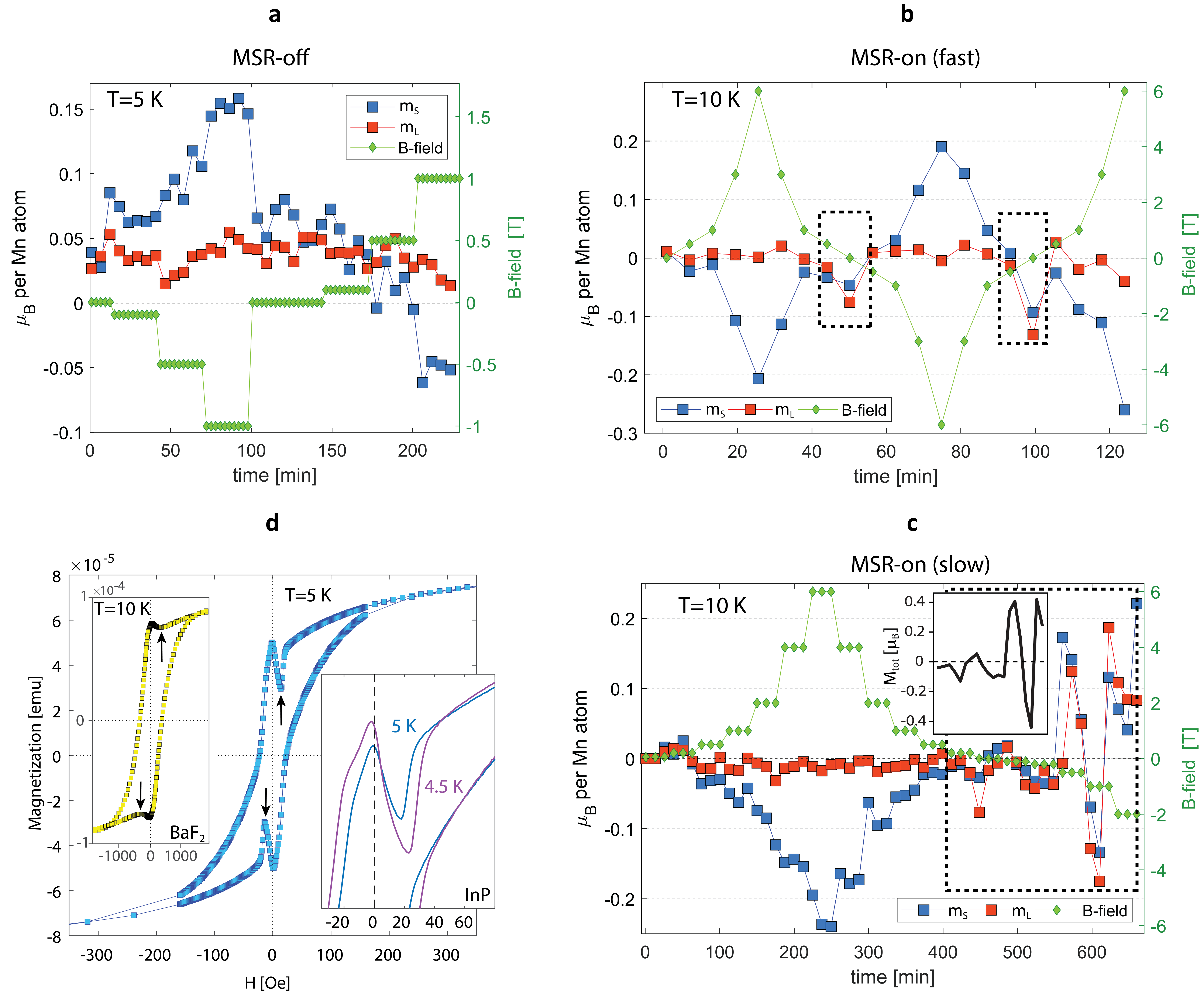}
    \caption{\textbf{Emergence of topological spin textures in XMCD and SQUID}. (a-c) Time series of spin (\SM) and orbital (\OM) moments showing their dependence on applied B-field under MSR-off (panel a) and MSR-on regimes measured with ``fast'' (panel b) and ``slow'' B-field step changes (panel c). Data in (a) are measured after zero-field cooling (ZFC), data in (b-c) after full warming/cooling cycle under applied field of 6\ Tesla (FH-FC). The inset in c shows the total moment time dependence $\mathrm{M_{tot}}$ from data inside the dashed frame. (d) Out-of-plane SQUID hysteresis for \GMTD\ grown on InP(111) (blue markers, right inset), and $\mathrm{BaF_2}$(111) substrates (left inset).}
    \label{F5}
\end{figure}

\begin{figure}[ht]
    \includegraphics[width=\textwidth]{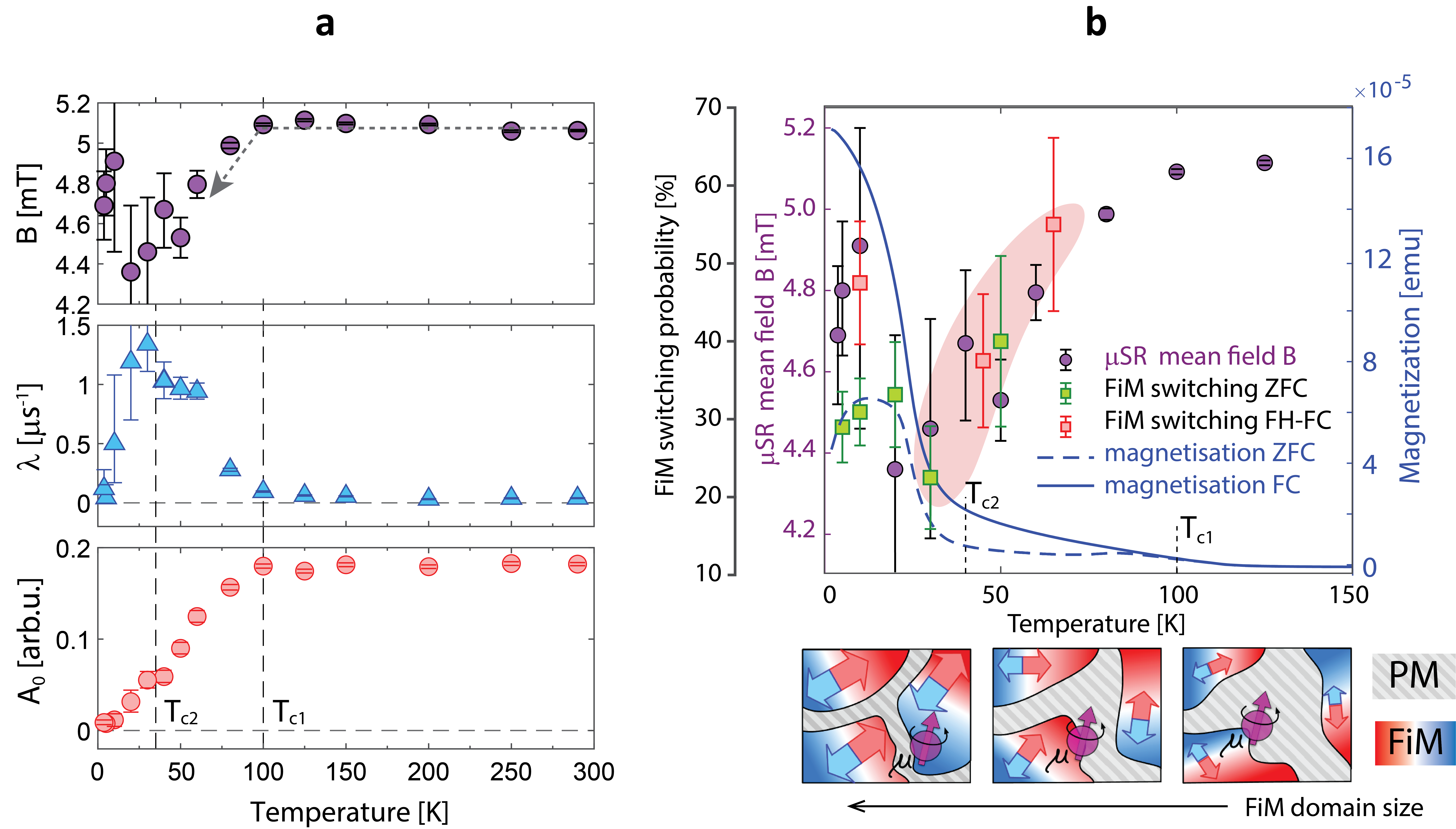}
    \caption{\textbf{Correlated spin glass state observed by LE-$\mu$SR, SQUID, and switching dynamics}. (a) Temperature-dependent average local field $B$ (top), muon damping rate $\lambda$ (middle), the initial amplitude $A_0$ (bottom), obtained from fits of the LE-$\mu$SR data. (b) Temperature dependent comparison of the $\mu$SR local mean field, bulk magnetization $M$ measured by SQUID magnetometry, and XMCD switching statistics for ZFC (green markers) and FH-FC (red markers) \cite{statNote}. The pictographs illustrate a muon stopping in FiM clusters embedded in a PM sea. As the clusters grow in size at lower temperatures, the average internal $B$-field shows a negative shift indicated by arrow in (a).}
    \label{F6}
\end{figure}

\clearpage
\newpage
\section*{Methods}

\setcounter{section}{0}
\renewcommand{\thesubsection}{\Roman{subsection}}  

\subsection{Sample preparation} \label{M_sprep}
We have grown 200 nm \GMTD\ thin films on InP(111) and BaF$\mathrm{_2}$(111) substrates by molecular beam epitaxy. The choice for 13\% \hbox{Mn-doping} is that for such a doping the Zeeman splitting is saturating \cite{JK_GMT}.  A protective stack of amorphous Te- and Se-capping layers with a total thickness of $\approx$20~nm was used to avoid surface oxidation and degradation. Before mounting into the XTREME beamline sample holder \cite{Piamonteze_XTREME_2012}, it was completely removed in a ultrahigh vacuum chamber by annealing the samples for 30-45~min at 250\degC; and recapped with $\approx$2~nm amorphous carbon by carbon-thread evaporation within the same vacuum chamber.

\subsection{Experimental methods} \label{M_expm}
The XMCD setup at the XTREME beamline \cite{Piamonteze_XTREME_2012} allows us to measure magnetic moments from the averaged orbital and spin contribution projected along the \hbox{out-of-plane} magnetization $\vec{M}$, which is collinear with the ferroelectric polarization $\vec{P}$ along the [111] direction \cite{JK_GMT}. Raw TEY and TFY signals measured with Keithley 6517A digital multimeters were normalized with the $\mathrm{I_0}$ photocurrent from the XTREME beamline refocussing mirror. All the signals were sampled from digital multimeter analog outputs, and further processed for the on-the-fly XMCD energy scans with a dedicated VME-system readout system by sampling/averaging the signals with 1~kHz and 100~Hz, respectively \cite{JK_OTF_AIP_2010}. The XMCD spectrum was obtained by taking the difference of ($\mu$+ - $\mu$-, where $\mu$+ and $\mu$- represent the XAS spectra measured by TEY or TFY and using the circular polarized X-rays with variable magnetic field up to 6~T parallel and anti-parallel to the beam direction. All data presented were measured during three independent beamtimes on fresh samples prepared in exactly the same way. The same samples were used for further NEXAFS (SuperXAS beamline at Swiss Light Source) and SQUID magnetometry. Special extra-large samples capped with Au were prepared for low energy muon spin relaxation experiment (LE-$\mu$SR) at the $\mu$E4 muon beamline at the Paul Scherrer Institut \cite{Prokscha2008NIPA}.  LE-$\mu$SR time spectra are measured with two positron detectors placed opposite to each other for the considered samples. Temperature dependent asymmetry spectra were obtained under an external magnetic field of 5~mT applied perpendicular to the initial muon spin polarization direction. Data indicate magnetically ordered islands with a net magnetic moment aligned with the applied field, embedded in a sea of paramagnetic environment. As these islands grow in size, their magnetic moment along the applied field increases and the strength of their dipolar field in the paramagnetic regions increases. This dipolar field is primarily opposite the direction of the applied field and hence produces the observed negative shift in B(T) as seen in Fig.~\ref{F6}a.

In order to achieve comparable local probe conditions with the XMCD photoelectron escape depth ($\approx$20~nm), the muon implantation energy was adjusted to set their stopping distribution just below the sample surface. Under magnetization, the asymmetry of the transverse field the muon precession decays. The main quantity determining the LE-$\mu$SR spectra in Fig.~\ref{F6}a is the muon spin precession amplitude decay from asymmetric emission of positrons, analyzed with \textit{Musrfit} program \cite{Musrfit_2012}. 

\subsection{Crystal structure and computational details}{\label{M_comp}

Using first-principles calculations within the density functional theory, we have investigated the ground state magnetic properties of bulk Ge$_{0.867}$Mn$_{0.133}$Te. The calculations were carried out using the multiple scattering KKR Green function method as implemented in the spin-polarized fully relativistic Korringa-Kohn-Rostoker (SPRKKR) code \cite{Minar_RPP}. The exchange and correlation effects were incorporated within the LDA framework \cite{Vosko_1980}. Brillouin zone integrations were performed on a 39$\times$39$\times$39 dense mesh of $k$-points. The angular momentum expansion up to l$_{\rm max}$=~4 has been used for each atom. The energy convergence criterion and coherent potential approximation (CPA) tolerance has been set to 10$^{-7}$ Ry. The potential is constructed in full potential geometry. The influence of chemical disorder in Ge$_{0.867}$Mn$_{0.133}$Te can be estimated by 	means of the CPA alloy theory \cite{Soven67,Taylor67} implemented in the SPR-KKR method. 

We consider two types of \Mni\ atoms denoted as $\mathrm{Mn_2}$ and  $\mathrm{Mn_3}$ in Fig.~\ref{F2}a, which occupy allowed Wyckoff positions within the \GMTD\ space group (see Table~\ref{T1}), derived from the experimental data as determined from the room temperature X-ray powder diffraction measurements \cite{Johnston_1961}.

\begin{table}[h] 
		\begin{center} 
			\begin{tabular}{ c} \\ \hline
	{{\bf Ge$_{0.867}$Mn$_{0.133}$Te}}  \\
	{space group: R3mR (No. 160), $a$=4.234\AA, {$\alpha$}=59.07$^{\circ}$}   
					\end{tabular} \\  
					\begin{tabular}{ c        c          c             } \\ \hline                 
		Wyckoff sites with atomic positions     &  \hspace{0.5 cm}  Atomic occupancies   \\\hline
		{ 1a (0.534,0.534,0.534)} & 0.911Ge + 0.089\Mns \\\hline
	    { 1a (0.0,0.0,0.0)} & Te \\\hline
	    { 1a (0.250, 0.250, 0.250)} & 0.978Es + 0.022Mn$_{\rm 2}$ \\\hline
	    { 1a (0.764, 0.764, 0.764)} & 0.978Es + 0.022Mn$_{\rm 3}$ \\\hline 
	\end{tabular} \\
	\end{center}
			\caption{Ge$_{0.867}$Mn$_{0.133}$Te R3mR (No. 160) space group crystal structure with atomic positions and occupancies. $Es$ represents empty spheres used in the CPA formalism, \Mni\ with two atomic positions (Mn$_2$,Mn$_3$) occupying allowed Wyckoff positions of R3mR space group. The atomic distribution of Mn atoms as substitutional (Mn$_2$; 67{\%}) and interstitial (Mn$_2$,Mn$_3$; 33{\%}) were obtained from NEXAFS measurements summarised in Extended Fig.~\ref{S1}c.}

\label{T1}
\end{table} 


\begin{table}[h!]
  \begin{center}
    
    \begin{tabular}{c c c c| c c c}
      
      \hline
      \multicolumn{4}{c|}{\shortstack{\vspace{0.1em} \\CPA theory \\{$\mu_B$/atom}}} 
      & 
      \multicolumn{3}{c}{\shortstack{XMCD (\Mns+\Mni)\\ {$\mu_B$/atom}}} \\ 
      \hline
         & \Mns & \Mni & $\Sigma\mathrm{Mn_{s,i}}$ & MSR-off 0 T &  MSR-on 0 T & MSR-on 6 T\\
      \hline
      \SM\ & 4.209  & -2.112 & 2.097 & 0.058$\pm$0.010 & $\approx$-0.089 & $\approx$0.238\\
      \OM\ & 0.047  & -0.079 & -0.032 & 0.040$\pm$0.004 & $\approx$-0.014 & $\approx$-0.012\\
      \hline
    \end{tabular}
  \end{center}
 \caption{Spin (\SM) and orbital (\OM) magnetic moments CPA theory vs. XMCD from \GMTD/InP[111], \SM\ and \OM\ momenta were calculated along the \GMTD[111] easy magnetization axis. }
 \label{T2}
\end{table}

In addition we also performed 2$\times$2$\times$2 super-cell QuantumESPRESSO pseudopotential relaxation calculations with the Mn atomic positions summarized in Table~\ref{T1}.  The results confirmed that AFM is the energetically most stable configuration. The structural models from the super-cell calculations were used to model XAS and XMCD calculations, which in turn where consistent with our CPA model summarized in Fig.~\ref{F2}d. Moreover, the structural relaxations around the Mn atoms showed negligible impact on the magnetic structure and shape of the XAS spectra.

Finally, we also performed multiplet calculations  based on localized Mn$^{2+}$ $3d$ with parameters reported by Sato \textit{et al.} \cite{Sato_JPRPh_2005}. The comparison  with measured \GMTD\ XAS spectra in Extended Fig.~\ref{S1}c and Fig.~\ref{S2}a confirms that our XAS and XMCD spectra feature localized Mn$^{2+}$ $3d$ states. Therefore, for the XMCD sum rule calculations we considered 5 holes in the Mn unoccupied $d$-states with a correction factor of 1.47 for compensating the $jj$-mixing \cite{Goering_2005}. However, as reported by Sato \textit{et al.} \cite{Sato_JPRPh_2005}, multiplet calculations suffer from intensity deviations between the calculations and experiments. For a better comparison with experimental data we thus relied on CPA calculations.

Because the CPA enables us to address spectral contributions from individual Mn-atoms, we tentatively modelled the resulting magnetic dichroism as a simple sum of all Mn contributions in saturation, as shown in Extended Fig.~\ref{S2}b.  When reducing the \Mns\ spectral contribution a factor 3, we find that the resulting XMCD signal (dashed line in panel d) reproduce the measured data very well, whereas multiplet calculations with $\mathrm{Mn_s^{2+}}$ atoms fail to reproduce the exact shape of the dichroism effect and lead to a spectral shift indicated by the black arrow (see Extended Fig.~\ref{S2}d). The factor 3 reduction of the \Mns\ spectral contribution is rationalized by observing a \textit{significant decrease} in \Mns\ dichroism between state \textcircled{\small{1}} right after the FH-FC@6T at 10~K, and \textcircled{\small{2}} recorded later during \hbox{MSR-on} conditions (vertical arrow in Extended Fig.~\ref{S2}b). With every FiM switching event at low temperatures some of the \Mns\ becomes part of the paramagnetic spin glass background; and when around one third of the original dichroism effect is left, this process appears to saturate (Extended Fig.~\ref{S2}c). However, on increasing the temperature from 10$\rightarrow$45-65~K, we observe a \textit{significant increase} in the \Mns\ dichroism, which we attribute to self-induced magnetization stimulated by MSR along with the paramagnetic spin glass background as schematically depicted in Extended Fig.~\ref{S4}b}.  

\subsection{Magnetic ground state calculations} {\label{M_Ham}}

Figure \ref{F2}a is a cartoon descriptions of various Mn-Mn exchange interactions between individual $\vec {S_1}$, $\vec {S_2}$ spin states considered in our model Hamiltonian where Mn impurities act at the same time as random magnetic moments and as acceptors producing the carriers. For the \GMTD\ magnetic ground state we consider the general spin Hamiltonian in the following form:

\begin{equation}
H = -\frac{1}{2}\sum_{ij}J_{,ij} \vec{S_i}\cdot\vec{S_j} + \sum_{ij}\vec{D_{i,j}}\cdot(\vec{S_i} \times \vec{S_j})+ \sum_{i}K_i\vec{S_i}^{2}
\end{equation}

\noindent The first term represents the Heisenberg exchange energy with $\vec{S_i}$ and $\vec{S_j}$ unit vectors having directions corresponding to local magnetic moments on sites $i$ and $j$. $J_{i,j}>0$ and $J_{i,j}<0$ prefer FM and AFM spin configurations, respectively. The second term originates from the antisymmetric part of the interaction matrix, also termed as antisymmetric exchange or Dzyaloshinskii-Moriya interaction (DMI). Finally the $K_i$ term in the final term accounts for the magnetocrystalline anisotropy energy (MAE).

The isotropic $J_{i,j}$ exchange between localized  $\vec {S_1}$, $\vec {S_2}$ spin states depends on the distance between the next-nearest Mn dopants. In contrast to earlier theoretical \GMT\ studies considering only \Mns\ atoms \cite{Fukushima_2014}, another type of interaction is coming from the \mbox{$\sum_{i,j}\vec{D_{i,j}}\cdot(\vec{S_i} \times \vec{S_j})$} \hbox{Dzyaloshinskii-Moriya} interaction obtained by sampling the magnetization deflection between [111] and [$\overline{111}$] easy magnetization axis directions. Below we discuss the contributions of individual terms to the \GMTD\ magnetic ground state.

\subsubsection*{Isotropic Exchange Interactions}

The magnetic exchange coupling parameters ($J_{i,j}$) are based on the real space approach by using the theory proposed by Liechtenstein \textit{et al.}\cite{Liechtenstein87}. This approach employs the ``magnetic force theorem'' to determine $J_{i,j}$ by assessing the total energy change related to an infinitesimal rotation of the magnetic moments located at the atomic sites $i$ and $j$. The energy change can be related to the exchange coupling parameters $J_{i,j}$ as:
	\begin{equation}
	J^{}_{i,j} =    \frac{1}{4\pi}\int^{E_F}dEImTr_L{\{}{\Delta^{}_{i}\tau^{i,j}_{\uparrow}\Delta^{}_{j}\tau^{j,i}_{\downarrow}}{\}} 
	\end{equation}
where $\tau$ is the scattering path operator, $\Delta_{i}$ is the difference in the inverse single site scattering $\it{t}$ matrices for spin up and spin down electrons, and Tr$_L$ is the trace of scattering matrices over the orbital indices $\it{L={(l,m)}}$.

The $J_{i,j}$ calculations are performed within a cluster of radius 3$a$, where $a$ is the lattice parameter. We neglect all interactions involving Ge and Te atoms and consider only those between Mn atoms which are found to host significant localized magnetic moments/atom. Fig.~\ref{F2}a of the main text shows three different types of Mn atoms inside the Ge$_{0.867}$Mn$_{0.133}$Te primitive unit cell, classified as substitutional \Mns\ and interstitial \Mni\ atoms, respectively. 

The physical mechanism behind the magnetic ground state has been attributed to $pd$-exchange coupling \cite{Fukushima_2014,Belhadji_2007}, combined with \hbox{$d$-$d$} magnetic interactions \cite{Sato_JPRPh_2005}. Consistently with our theoretical predictions, the $pd$-exchange coupling is relatively weak, but long ranged. Moreover, for \GMT\ the $pd$-exchange constant is negative \cite{Sato_JPRPh_2005}, stemming from AFM exchange interaction between $\mathrm{Mn^{2+}}$ states. In agreement with our \GMT\ magnetic ground state description, this negative term naturally materialize between \Mni-\Mns\, which is further confirming that \GMT\ is a ferrimagnetic rather than a ferromagnetic DMS \cite{Fukuma_2005,Sato_JPRPh_2005, Fukuma_2006, Fukuma_2008, Fukushima_2014}.

\subsubsection*{Dzyaloshinskii-Moriya interaction}

The DMI is a chiral exchange interaction between localized spins and has a net contribution only in systems without structural inversion symmetry and with the presence of an indirect or long-range exchange interaction. \GMTD\ satisfies these two conditions, which is also evident from the isotropic exchange interactions described in the previous section. The origin of DMI is found in the spin-orbit coupling (SOC) which acts as a perturbation on localized orbital states. Given two neighboring spins $\vec{S_1}$ and $\vec{S_2}$, the DMI energy can be described by \mbox{-$\vec{D_{1,2}}\cdot(\vec{S_1} \times \vec{S_2})$}. Therefore, the direction of the relative $\vec{S_1}$ and $\vec{S_2}$ rotations can be clockwise or counter clockwise, providing the information about helicity in the case of spin spirals. This expression is part of a generalized exchange interaction where the DMI term  is related to the exchange constant $J$ of the direct Heisenberg exchange $-J(\vec{S_i}\cdot\vec{S_j})$. However,  contrary to the latter which favors collinear alignment, the DMI promotes an orthogonal arrangement between $\vec{S_i}$ and $\vec{S_j}$, with a chirality imposed by the direction of $\vec{D_{i,j}}$. The resulting spin helicity is uniform, meaning that clockwise or counterclockwise rotation of spins are energetically identical. Data in Fig.~\ref{F2}c show all the components of $\vec{D_{i,j}}$ (i.e. $D_{i,j,x}$,$D_{i,j,y}$,$D_{i,j,z}$) interaction, showing robust exchange up to a cluster size of 3.0$a$. The average DMI energy is comparable to the $J_{i,j}$ exchange energy for larger distances; and a factor 4 higher than the magnetic anisotropy energy (MAE) discussed below.

\subsubsection*{Magnetic Anisotropy Energy}

The magnetic ground state of FiM exchange energy (grey markers in Fig.~\ref{F3}a) has been explored by determining the total energy  corresponding to the relative spin orientation of $\mathrm{Mn_2}$ and $\mathrm{Mn_3}$ spins with respect to the substitutional $\mathrm{Mn_1}$ spins. The $\mathrm{Mn_1}$ spin was frozen along the [111] quantization axis and the $\mathrm{Mn_2}$, $\mathrm{Mn_3}$ spins were rotated along the out-of-plane [111] direction starting from {$\theta$}=0, which corresponds to the FM state with all spins pointing in the same direction. However, the self consistency convergence in calculations can be achieved only for a limited $\theta$ range between 40$^{\circ}$-320$^{\circ}$.  Nevertheless, the total energy of the system reaches a minimum with $\mathrm{Mn_1}$ spins set to 0$^{\circ}$ and $\mathrm{Mn_{2,3}}$ spins to 180$^{\circ}$, which establishes the FiM order. The red markers in Fig.~\ref{F3}a show the MAE energy variation as a function of magnetization angle $\theta$, evaluated as a difference between the fully relativistic total energies calculated for quantization axes [111] and axis orthogonal to  $\bar{[111]}$. Data confirm that the easy magnetization axis is along the [111] or [$\bar{1}\bar{1}\bar{1}$] directions. The calculated MAE of the Ge$_{0.867}$Mn$_{0.133}$Te is 0.101 meV/f.u. and represents the energy barrier between the easy and hard magnetic axes. Finally, the energy landscape obtained by combination of MAE and FiM ground state was used to model the FiM double-well potential as depicted in Fig.~\ref{F3}a.  The MSR-driven FiM switching contracts the states into one of the two wells. The bistable potential depicted can be expressed as \mbox{$U(x)=-\frac{1}{2}ax^2+\frac{1}{4}bx^4$} with minima located at $\pm$90$^{\circ}$ and barrier height $\Delta V=a^2/4b$.

\clearpage
\newpage

\setcounter{figure}{0} 
\section*{Extended data}

\begin{figure}[ht!]
    \includegraphics[width=\textwidth]{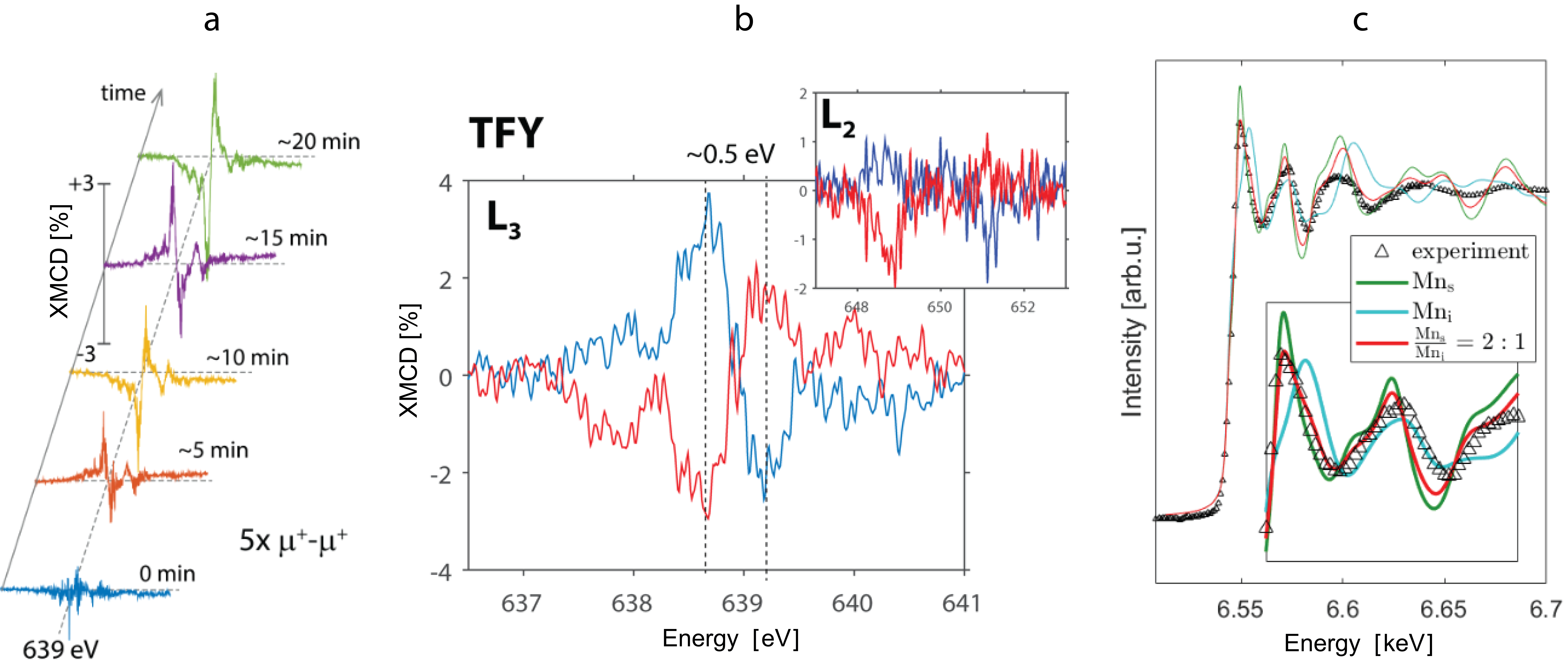}
    \caption{(a) Stacked plot of \MnL\ XMCD spectra measured with same circular light polarization $\mu^+$ under \hbox{MSR-on} regime. A change in x-ray helicity relative to the fixed magnetization direction is equivalent to a change in the magnetization direction relative to the fixed helicity. (b) Spontaneous switching in consecutive XMCD datasets recorded in total fluorescence yield (TFY) at T=5\ K and B=0\ T, zoomed at the \MnL\ and $\mathrm{Mn_{L2}}$ absorption edges. (c) NEXAFS Mn-K edge data fits. Panel inset shows the best fit obtained with \Mns:\Mni\ occupancy close to 2:1.}
    \label{S1}
\end{figure}

\begin{figure}
    \includegraphics[width=\textwidth]{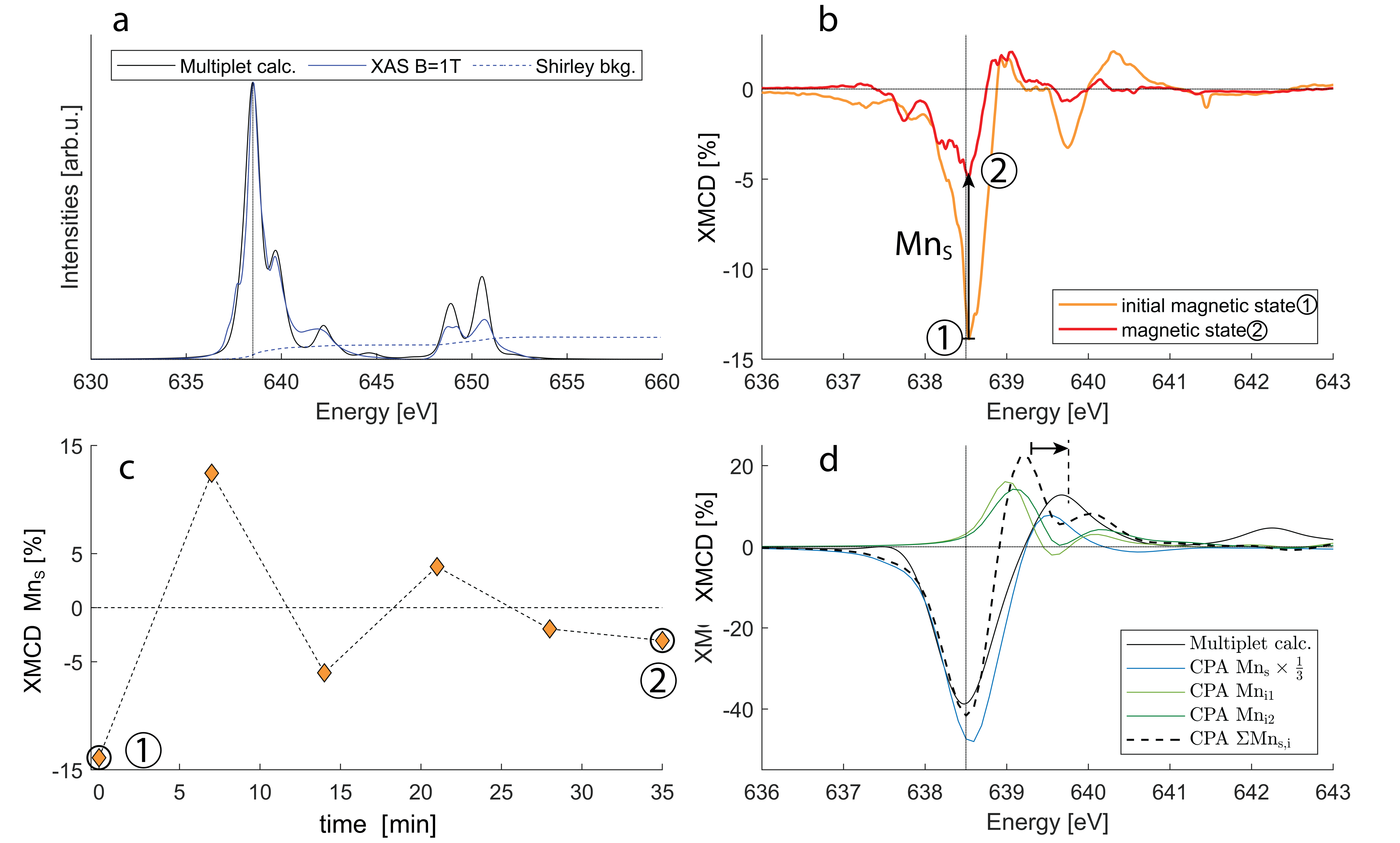}
    \caption{(a) Background subtracted XAS spectra measured at 6\ T (10\ K), compared with multiplet calculations. (b) \MnL\ XMCD measured after ZFC at 1\ T (blue trace), compared with FH-FC spectra in \Mns\ state \textcircled{\small{1}} and \textcircled{\small{2}}, respectively. (c) Time dependence of the dichroism effects under MSR-on regime at \MnL-edge between the initial magnetic state \textcircled{\small{1}} after FH-FC and magnetic state \textcircled{\small{2}}. (d) Calculated XMCD spectra using CPA and multiplet calculations. For CPA spectral contributions from individual Mn atoms are indicated. CPA is in good agreement with the experimental data, whereas the multiplet calculations show a clear shift in the spectral features as indicated by the horizontal arrow.}
    \label{S2}
\end{figure}

\begin{figure}
    \includegraphics[width=0.9\textwidth]{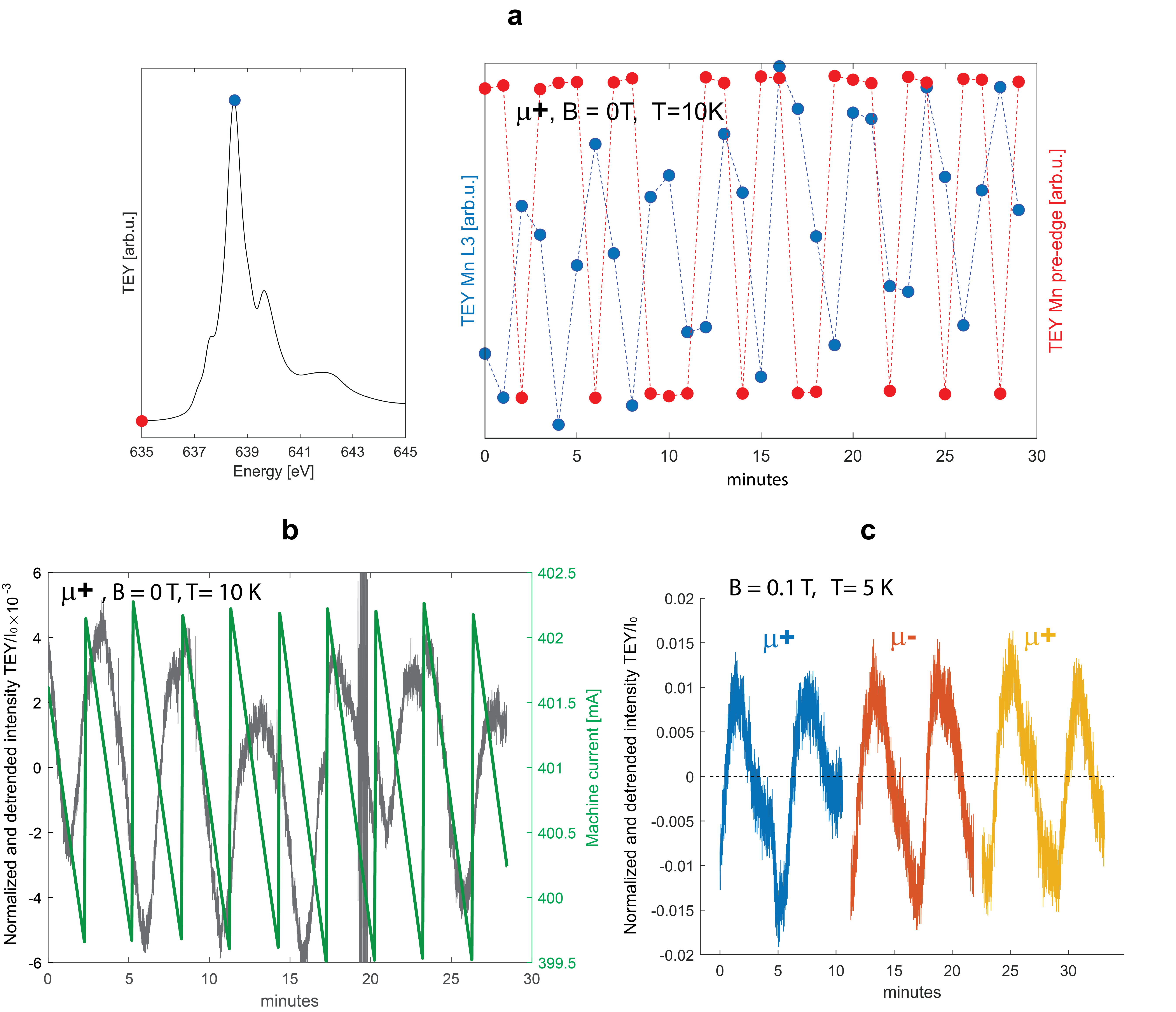}
    \caption{(a) TEY time survey of \MnL\ pre-edge (red markers) and edge (blue markers) obtained during XMCD acquisition. (b) TEY time survey at  $h\nu=638.8$eV (\MnL\ edge) overlaid with the machine top-up current. (c) Series of TEY time surveys at $h\nu=638.8$eV under applied B-field and different x-ray polarizations.}
    \label{S3}
\end{figure}

\begin{figure}
    \includegraphics[width=\textwidth]{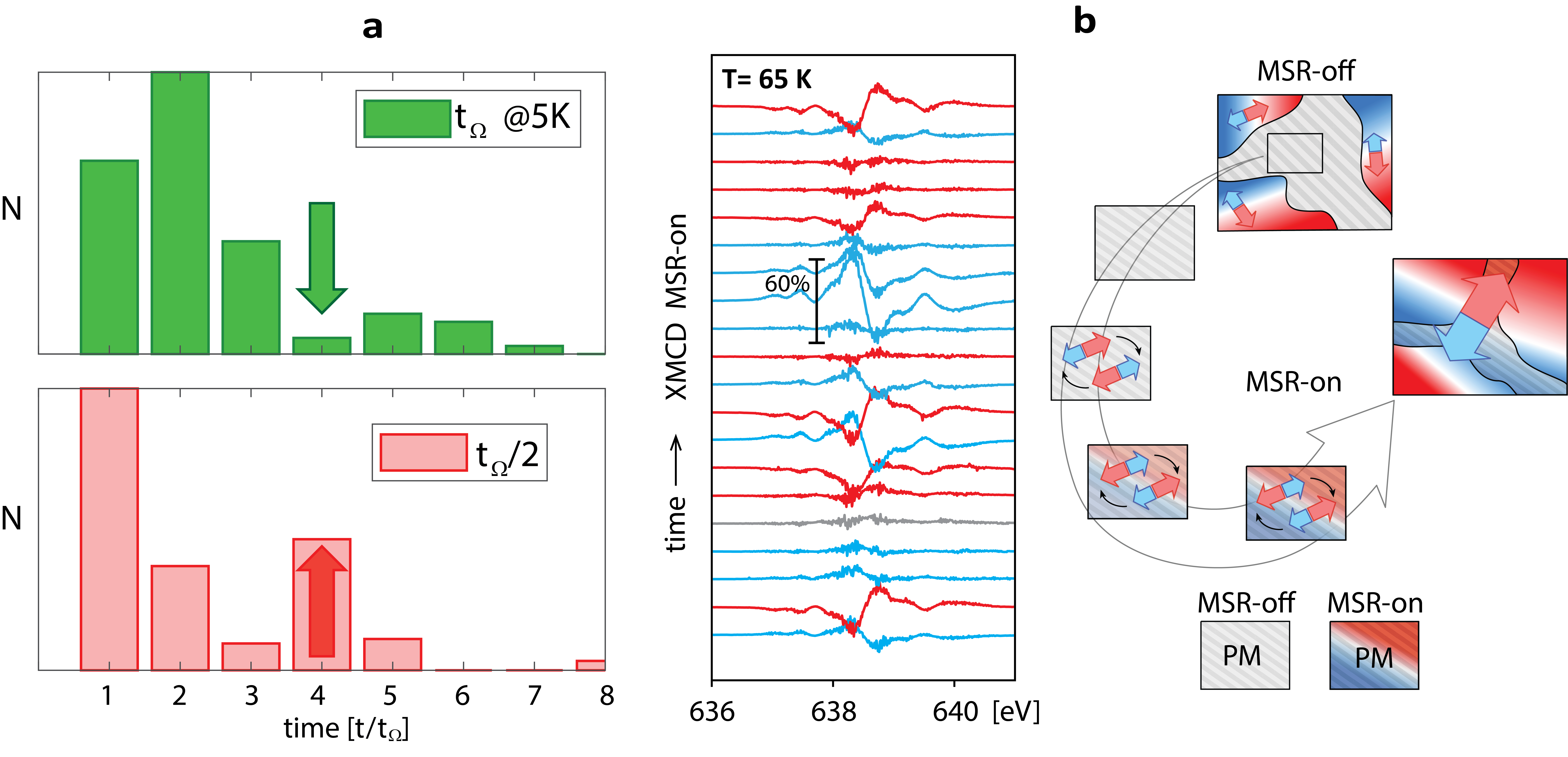}
    \caption{(a) (top) Magnetostochastic switching residence time distribution $N(t)$ for a data set measured at 5\ K, ZFC and $t_{\Omega}$ periodic drive obtained by measuring full XMCD spectra covering both \MnL+$\mathrm{Mn_{L2}}$; compared with all data measured after FH-FC for 5, 45 and 65\ K (red markers in Fig.~\ref{F6}b) where the switching statistics was evaluated from shorter energy scans across the \MnL\ only, i.e. by halving the periodic drive ($\mathrm{\frac{1}{2}t_{\Omega}}$). The arrows highlight the $N(t)$ phase shift due to different stochastic resonance conditions \cite{Gammaitoni_RMP_1998}.  (b) Series of \MnL\ XMCD spectra measured at 65\ K, with maximum dichroism effect close to 60\%. The pictographs illustrate enhancement of the FiM order (red-blue arrows) spreading across the magnetic clusters along with self-induced magnetization of the PM sea under MSR FiM switching.}
    \label{S4}
\end{figure}

\end{document}